\documentstyle[12pt,qqaalart]{article}

\author{Jos\'e Antonio Belinch\'on\thanks{Interuniversity Group of Dimensional Analysis. Dept. of Physics  ETS Arquitectura Madrid :E-mail: jabelinchon@usa.net}\\
C/ Conil N 5\\
Pozuelo de Alarc\'on\\
Madrid 28224\\
Spain}
\title{Some FRW viscous models with $G$, $c$ and $\Lambda $ variables
}
\input tcilatex

\QQQ{Language}{
American English
}

\begin{document}

\maketitle
\begin{abstract}
We consider several viscous models with metrics FRW ($k=0)$ but with
variable $G,c$ and $\Lambda $ . We find trivially a set of solutions through
Dimensional Analysis.
\end{abstract}

\section{\bf Introduction.}

Recently have been studied several models with metrics FRW where the
''constants'' $G$ and $\Lambda $ (\cite{BA}) are considered as dependent
functions on time $t$ $(\cite{A})$. Most recently this type of models have
been generalized by Arbab (\cite{AB}) who considers a viscous fluid. Other
authors (\cite{BA}) study models with $c$ and $G$ variables for perfect
fluids. In this paper we want to calculate as vary these ''constants'' $G,c$
and $\Lambda $ within the models FRW with a viscous fluid. We want to
emphasize as the use of the Dimensional Analysis (D.A.) permits us to find
in a trivial way a set of solutions to this type of models (but with $k=0),$
taking into account the conservation principle $(divT_{ij}=0)$, since the
type of differential equations that go emerging can be (are) cumbersome. We
have not wanted to make a meticulous study of the solutions obtained for
each $(\omega ,n)$ (we let this paragraph for a subsequent paper) but to
show the obtained results.

The paper is organized as follows: In the second paragraph are made some
small considerations on the followed dimensional method (address to reader
to the classic literature on the topic (\cite{B})). In the third paragraph
we make use of the Dimensional Analysis ( Pi theorem) to obtain a solution
to the principal quantities that appear in the model and finally in the
fourth paragraph we end with a short exposition of the obtained cases.

\section{\bf The model.}

The modified field equations are: 
\begin{equation}
\label{e0}R_{ij}-\frac 12g_{ij}R-\Lambda (t)g_{ij}=\frac{8\pi G(t)}{c^4(t)}%
T_{ij} 
\end{equation}
and we impose that%
$$
div(T_{ij})=0 
$$
where $\Lambda (t)$ represents the ''cosmological constant''. The basic
ingredients of the model are:

The line element is defined by:%
$$
ds^2=-c^2dt^2+f^2(t)\left[ \frac{dr^2}{1-kr^2}+r^2\left( d\theta ^2+\sin
{}^2\theta d\phi ^2\right) \right] 
$$
here only we will consider the case $k=0.$ The energy-momentum tensor is
defined by:%
$$
T_{ij}=(\rho +p)u_iu_j-pg_{ij}\qquad \qquad p=\omega \rho 
$$
The developed equations of the field are: 
\begin{equation}
\label{e1}2\frac{f\,^{\prime \prime }}{f\,}+\frac{(f\,^{\prime })^2}{f\,^2}= 
\frac{8\pi G(t)}{c^2(t)}p+c^2(t)\Lambda (t)\ \ 
\end{equation}

\begin{equation}
\label{e2}\frac{(f\,^{\prime })^2}{f\,^2}=\frac{8\pi G(t)}{3\,c^2(t)}\rho
+c^2(t)\Lambda (t)\qquad \quad \ 
\end{equation}

\begin{equation}
\label{e3}div(T_{ij})=0\text{ }\Leftrightarrow \rho ^{\prime }+3(\omega
+1)\rho \frac{f^{\prime }}f=0 
\end{equation}
integrating the equation (\ref{e3}) we obtain the following equation. 
\begin{equation}
\label{e4}\rho =A_\omega f^{-3(\omega +1)} 
\end{equation}
where $f$ represents the scale factor that appears in the metrics and $%
A_\omega $ is the constant of integration that depends on the equation of
state that is imposed. The effect of the viscosity in the equations is shown
replacing $p$ by $p-3\eta H$ ($H=f^{\prime }/f)$ (\cite{AB} and \cite{BAR}
for details). where: 
\begin{equation}
\label{n1}\eta =\eta _0\rho ^n 
\end{equation}
This last equation (of state) in our opinion does not verify the dimensional
homogeneity principle, by this reason we have the changed it by: 
\begin{equation}
\label{n2}\eta =k_n\rho ^n 
\end{equation}
where the constant $k_n$ causes that the above equation will be
dimensionally homogeneous for any value of $n.$ The dimensional analysis
that we apply needs to make the following distinctions. We need to know
beforehand the set of fundamental quantities, in this case it is solely the
cosmic time $t$ as is deduced of the homogeneity and isotropy of the model
and to distinguish the set of constants, universal and unavoidable or
characteristic, in this case there are no universal constants since are all
functions on $t$ $(G(t),c(t),\Lambda (t))$ $(G(t),c(t),,$that is to say we
do not consider them, and the two only constants that appear are
respectively the constant of integration $A_\omega $ that depending on the
equation of state that is imposed will have different dimensions and
physical meaning and the constant $k_n$ that it will depend on the equation
of state that we impose for $\eta .$

In a previous paper (\cite{T}) was calculated the dimensional base of this
type of models, being this $B=\left\{ L,M,T,\theta \right\} $ where $\theta $
represents the dimension of the temperature. The corresponding dimensions of
each magnitude (with respect to this base) are:%
$$
\left[ t\right] =T\quad \left[ A_\omega \right] =L^{2+3\omega }MT^{-2}\qquad
\left[ k_n\right] =L^{n-1}M^{1-n}T^{2n-1} 
$$
All the magnitude that we go to calculate we will make it exclusively in
function of the cosmic time $t$ and of the constant unavoidable $k_n$ and $%
A_\omega $ with respect to a dimensional base $B=\left\{ L,M,T,\theta
\right\} .$

\section{{\bf {Solutions through A.D.}}}

We go to calculate through dimensional analysis D.A. the variation of $G(t)$
in function of $t$, the speed of the light $c(t),$ energy density $\rho
@(t), $ the radius of the universe $f(t),$ the temperature $\theta (t)$, and
finally $\Lambda (t)@.$

\subsection{{\bf Calculation of }${\bf G(t)}$}

As have indicated above, we go to accomplish the calculation applying the Pi
theorem. The magnitude that we consider are: $G=G(t,k_n,A_\omega ).B=\left\{
L,M,T,\theta \right\} .$ We know that $\left[ G\right] =L^3M^{-1}T^{-2}$%
$$
\left( 
\begin{array}{rrrrr}
& G & t & k_n & A_\omega \\ 
L & 3 & 0 & n-1 & 2+3\omega \\ 
M & -1 & 0 & 1-n & 1 \\ 
T & -2 & 1 & 2n-1 & -2 
\end{array}
\right) 
$$
\begin{equation}
\label{r1}G\propto A_\omega ^{-1+\frac{3\omega +5}{3(\omega +1)}}k_n^{\frac{%
3\omega +5}{3(\omega +1)(n-1)}}t^{-4-\frac{3\omega +5}{3(\omega +1)(n-1)}} 
\end{equation}

\subsection{\bf Calculation of $c(t)$}

$c(t)=c(t,k_n,A_\omega )\Longrightarrow $

{\bf 
$$
c(t)\propto A_\omega ^{\frac 1{3(\omega +1)}}k_n^{\frac 1{3(\omega
+1)(n-1)}}t^{-1-\left[ \frac 1{3(\omega +1)(n-1)}\right] } 
$$
}

\subsection{\bf Calculation of energy density $\rho (t)$}

$\rho =\rho (t,k_n,A_\omega )$ with respect to the dimensional base $B$%
\begin{equation}
\label{r3}\rho \propto k_n^{\frac 1{1-n}}t^{\frac 1{n-1}} 
\end{equation}

\subsection{\bf Calculation of the radius of the universe $f(t).$}

$f=f(t,k_n,A_\omega )\Longrightarrow $%
\begin{equation}
\label{r4}f\propto A_\omega ^{\frac 1{3(\omega +1)}}k_n^{\frac 1{3(\omega
+1)(n-1)}}t^{\frac{-1}{3(\omega +1)(n-1)}} 
\end{equation}
We can observe that:%
$$
q=-\frac{f^{\prime \prime }f}{\left( f^{\prime }\right) ^2}=-\left[ 3(\omega
+1)(n-1)+1\right] 
$$
$$
H=\frac{f^{\prime }}f=-\left( \frac 1{3(\omega +1)(n-1)}\right) \frac 1t 
$$
$$
d_H=ct\lim _{t_0\rightarrow 0}\int_{t_0}^t\frac{dt^{\prime }}{f(t^{\prime })}%
=\infty 
$$
Thus, the model has no horizon because $d_H$ diverges for $t_0\rightarrow 0.$%
( depending of the model obviously)

\subsection{\bf Calculation of the temperature $\theta (t).$}

$\theta =\theta (t,k_n,A_{\omega ,}k_B)$ where $k_{B\text{ }}$ is the
Bolztmann constant $\Longrightarrow $%
\begin{equation}
\label{r5}k_B\theta \propto A_\omega ^{\frac 1{3(\omega +1)}}k_n^{\frac{%
-\omega }{(\omega +1)(n-1)}}t^{-2+\frac \omega {(\omega +1)(n-1)}} 
\end{equation}

\subsection{\bf Calculation of the cosmological constant: $\Lambda (t).$}

$\Lambda =\Lambda (t,k_n,A_\omega )$%
\begin{equation}
\label{r8}\Lambda \propto A_\omega ^{\frac{-2}{3(\omega +1)}}k_n^{\frac{-2}{%
3(\omega +1)(n-1)}}t^{\frac 2{3(\omega +1)(n-1)}} 
\end{equation}

\section{\bf Different cases.}

All the following cases have been studied by Arbab (\cite{AB}) and can be
without difficulty calculated the rest of the cases.

\subsection{$n=0$ y $\omega =0$}

$$
G\propto A_\omega ^{2/3}k_n^{-5/3}t^{-7/3}\quad \qquad c\propto A_\omega
^{1/3}k_n^{-1/3}t^{-2/3}\qquad \Lambda \propto A_\omega
^{-2/3}k_n^{2/3}t^{-2/3} 
$$
$$
\rho \propto k_nt^{-1}\qquad \qquad k_B\theta \propto A_\omega
^{1/3}k_n^0t^{-2} 
$$
$$
f\propto A_\omega ^{1/3}k_n^{-1/3}t^{1/3}\qquad q=2 
$$

\subsection{$n=1/2$ y $\omega =1/3$}

$$
G\propto A_\omega ^{1/2}k_n^{-3}t^{-1}\quad \qquad c\propto A_\omega
^{1/4}k_n^{-1/2}t^{-1/2}\quad \qquad \Lambda \propto A_\omega
^{-1/2}k_nt^{-1} 
$$
$$
\rho \propto k_n^2t^{-2}\quad \quad \quad k_B\theta \propto A_\omega
^{1/4}k_n^{-1/2}t^{-3/2} 
$$
$$
f\propto A_\omega ^{1/4}k_n^{-1/2}t^{1/2}\qquad q=1 
$$

\subsection{$n=3/4$ y $\omega =1/3$}

$$
G\propto A_\omega ^{1/2}k_n^{-6}t^2\quad \qquad c\propto A_\omega
^{1/4}k_n^{-1}t^0=const.\quad \qquad \Lambda \propto A_\omega
^{-1/2}k_n^2t^{-2} 
$$
$$
\rho \propto k_n^4t^{-4}\quad \quad \quad k_B\theta \propto A_\omega
^{1/4}k_nt^{-1} 
$$
$$
f\propto A_\omega ^{1/4}k_n^{-1}t\qquad q=0 
$$

\subsection{$n=2/3$ y $\omega =0$}

$$
G\propto A_\omega ^{2/3}k_n^{-5}t\quad \qquad c\propto A_\omega
^{1/3}k_n^{-1}t^0=const.\quad \qquad \Lambda \propto A_\omega
^{-2/3}k_n^2t^{-2} 
$$
$$
\rho \propto k_n^3t^{-3}\quad \quad \quad k_B\theta \propto A_\omega
^{1/3}k_n^0t^{-2} 
$$
$$
f\propto A_\omega ^{1/3}k_n^{-1}t\qquad q=0 
$$

\section{\bf Conclusions}

We have solved through Dimensional Analysis a viscous FRW model with $k=0$
and taking into account the energy-momentum tensor and with $G,\Lambda $ and 
$c$ variables. We observe that the solutions that we have obtained coincide
with the already obtained by Arbab unless that in this case our model
envisages the variation on $c,$ generalizing thus the solutions. We let the
conclusions and study of each case for a subsequent paper.

\end{document}